\documentclass{aastex63}

\usepackage{multirow}
\usepackage{float}

\newcommand\logg{log\,$g$}
\newcommand\Teff{$T_{\rm eff}$}

\submitjournal{AJ}

\shorttitle{DA stars}
\shortauthors{Yang et al.}

\graphicspath{{./}{figures/}}

\begin{document}

\title{Estimating Atmospheric Parameters of DA White Dwarf Stars with Deep Learning}

\correspondingauthor{Jingkun Zhao}
\email{zjk@bao.ac.cn}

\author[0000-0001-7609-1947]{Yong Yang}
\affiliation{Key Laboratory of Optical Astronomy, National Astronomical Observatories, Chinese Academy of Sciences, Beijing 100101, China}
\affiliation{School of Astronomy and Space Science, University of Chinese Academy of Sciences, Beijing 100049, China},
\author{Jingkun Zhao}
\affiliation{Key Laboratory of Optical Astronomy, National Astronomical Observatories, Chinese Academy of Sciences, Beijing 100101, China},
\author{Jiajun Zhang}
\affiliation{Key Laboratory of Optical Astronomy, National Astronomical Observatories, Chinese Academy of Sciences, Beijing 100101, China}
\affiliation{School of Astronomy and Space Science, University of Chinese Academy of Sciences, Beijing 100049, China},
\author{Xianhao Ye}
\affiliation{Key Laboratory of Optical Astronomy, National Astronomical Observatories, Chinese Academy of Sciences, Beijing 100101, China}
\affiliation{School of Astronomy and Space Science, University of Chinese Academy of Sciences, Beijing 100049, China},
\author{Gang Zhao}
\affiliation{Key Laboratory of Optical Astronomy, National Astronomical Observatories, Chinese Academy of Sciences, Beijing 100101, China}
\affiliation{School of Astronomy and Space Science, University of Chinese Academy of Sciences, Beijing 100049, China}

\begin{abstract}

The determination of atmospheric parameters of white dwarf stars (WDs) is crucial for researches on them. Traditional methodology is to fit the model spectra to observed absorption lines and report the parameters with the lowest $\chi ^2$ error, which strongly relies on theoretical models that are not always publicly accessible. In this work, we construct a deep learning network to model-independently estimate \Teff\ and \logg\ of DA stars (DAs), corresponding to WDs with hydrogen dominated atmospheres. The network is directly trained and tested on the normalized flux pixels of full optical wavelength range of DAs spectroscopically confirmed in the Sloan Digital Sky Survey (SDSS). Experiments in test part yield that the root mean square error (RMSE) for \Teff\ and \logg\ approaches to 900 K and 0.1 dex, respectively. This technique is applicable for those DAs with \Teff\ from 5000 K to 40000 K and \logg\ from 7.0 dex to 9.0 dex. Furthermore, the applicability of this method is verified for the spectra with degraded resolution $\sim 200$. So it is also practical for the analysis of DAs that will be detected by the Chinese Space Station Telescope (CSST).

\end{abstract}

\section{Introduction} \label{intro}

White Dwarf stars (WDs) are the final stage of evolution of stars whose progenitor masses are below 8-10.5 $M_{\sun}$, depending on metallicity \citep{2015MNRAS.446.2599D}, which contain important information needed to comprehend stellar formation and evolution. Most of WDs, especially single ones (i.e. not in binary system), will exist for a very long time due to their slow cooling processes, which means the cool ones may record information from extremely early time. Therefore, researches on those WDs are also helpful for understanding certain history of the Milky Way \citep[e.g.][]{2014ApJ...791...92T}.

The number of spectroscopically identified WDs is increasing thanks to SDSS \citep{2000AJ....120.1579Y,2011AJ....142...72E,2017AJ....154...28B} and most of them are actually DA types. \citet{2004ApJ...607..426K} reported 1888 DAs out of 2551 WDs based on the first Data Release \citep[DR1;][]{2003AJ....126.2081A}. Using SDSS DR4 \citep{2006ApJS..162...38A}, \citet{2006ApJS..167...40E} presented 8000 DAs out of 9316 WDs. Based on SDSS DR7 \citep{2009ApJS..182..543A}, \citet{2013ApJS..204....5K} reported 19713 WDs, 12831 of which were DAs. Furthermore, 6887 DAs out of 8441 WDs were presented by \citet{2015MNRAS.446.4078K} using SDSS DR10 \citep{2014ApJS..211...17A}. Based on SDSS DR12 \citep{2015ApJS..219...12A}, \citet{2016MNRAS.455.3413K} reported nearly 2883 WDs, 1964 out of which were DAs. 15716 DAs out of 20088 WDs were presented by \citet{2019MNRAS.486.2169K} in SDSS DR14 \citep{2018ApJS..235...42A}. The  atmospheric parameters of these WDs were determined from theoretical spectral fitting. Stellar spectra contain much information about a star and it is exactly this point that allows us to explore relation between spectra and parameters of a star in different ways.

Artificial neural network (ANN) has long been applied to the determination of astrophysical parameters. \citet{1997MNRAS.292..157B} trained an ANN on a set of synthetic optical stellar spectra to get \Teff, \logg\ and [M/H]. Nowadays, due to the development of computing power and availability of large data sets, ANN especially deep learning method has been widely used in astronomy to predict physical parameters of stars. \citet{2018MNRAS.475.2978F} applied Convolutional Neural Networks (CNN) to the data set from the Apache Point Observatory Galactic Evolution Experiment \citep[APOGEE;][]{2017AJ....154...94M}. \citet{2019ApJ...887..193L} and \citet{2019PASP..131i4202Z} trained deep neural networks respectively to estimate parameters and abundances of main-sequence stars using spectra from APOGEE and the Large Sky Area Multi-Object Fiber Spectroscopic Telescope \citep[LAMOST;][]{2006ChJAA...6..265Z,2012RAA....12..723Z}, which both yielded excellent performance. \citet{2020MNRAS.497.2688C} presented a generative fitting pipeline that interpolates theoretical spectra with ANN to determine atmospheric labels of WDs. However, there are still few works using ANN to estimate parameters of WDs so far.

In this work, a deep learning network is constructed based on an architecture called Residual Network (ResNet) \citep{2015arXiv151203385H} and is directly trained on the spectra of DAs identified in SDSS DR7, DR10 and DR12 to estimate atmospheric parameters. In section \ref{dataset}, we introduce the reduction of data. In section \ref{method}, our method and details of training and test are described. Finally, the conclusion of this work is given in section \ref{conclusion}.

\section{Data Set} \label{dataset}

We collect all spectra of DAs spectroscopically confirmed in SDSS DR7, DR10 and DR12. Corresponding atmospheric parameters are obtained from the catalogs published in \citet{2013ApJS..204....5K} and \citet{2015MNRAS.446.4078K, 2016MNRAS.455.3413K}. To begin with, the data are filtered with following criteria: (a) single DAs, (b) signal to noise rate in g-band $S/N_g\geq 10$, and (c) relative errors of \Teff\ and \logg\ $\leq 10\%$, which yield 9159 objects. Then we apply 3D correction (ML2/$\alpha$=0.8) to \Teff\ and \logg\ according to the calibration from \citet{2013A26A...559A.104T}.

\begin{figure}[htbp]
    \plotone{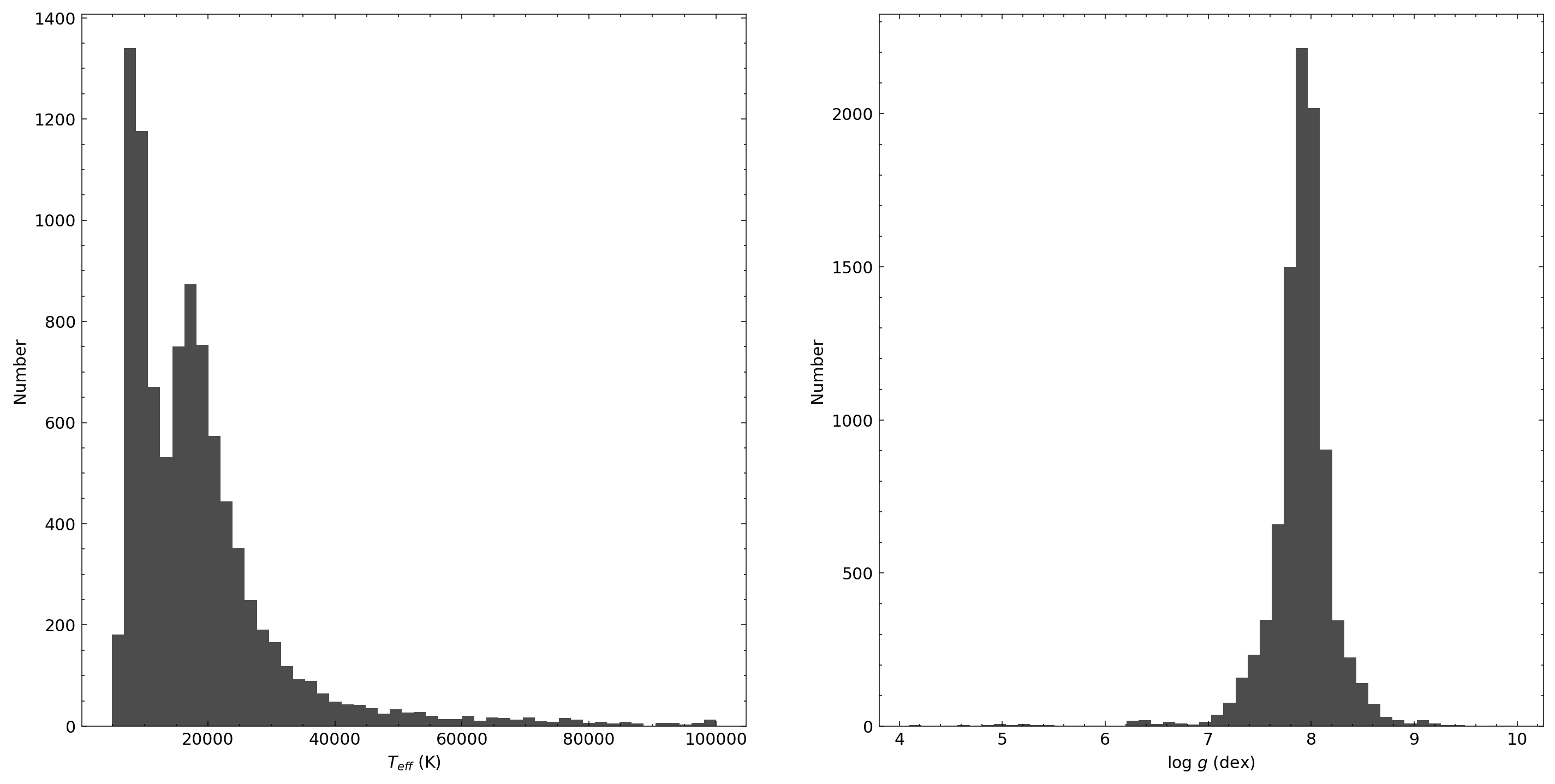}
    \caption{The distributions of \Teff\ and \logg\ of 9159 filtered DAs.  \label{fig:all_dist}}
\end{figure}

Figure \ref{fig:all_dist} shows the distributions of \Teff\ and \logg\ of filtered data. Based on this, we further select the objects that satisfy 5000 K $\leq$ \Teff\ $\leq$ 40000 K and 7.0 dex $\leq$ \logg\ $\leq$ 9.0 dex and finally get 8490 DAs. Next we preprocess them in two ways.

\subsection{Preprocessing on Original Spectra: Data Set 1} \label{pre1}

\begin{figure}[htbp]
    \plotone{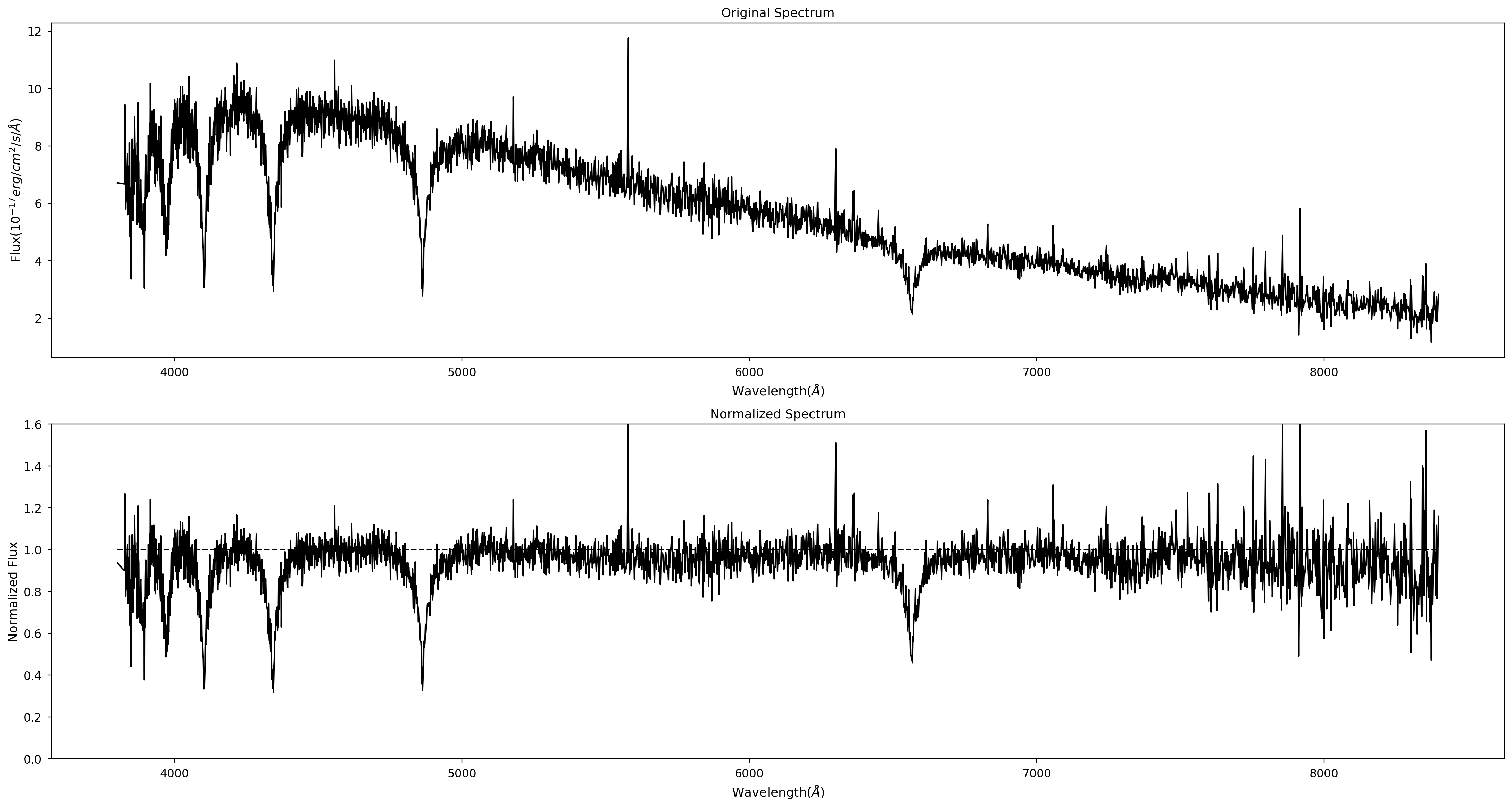}
    \caption{Observed and normalized spectrum of ``spSpec-54612-2922-626.fit'' in DR7.  \label{fig:original_wave}}
\end{figure}

We normalize original spectra of DAs with the script provided by \citet{2014A26A...569A.111B}. Firstly, a spectrum is re-sampled with an equivalent wavelength step. Next, a median and maximum filter with different window sizes are applied to find the continuum points. The median filter (here we set wavelength step = 30 in the script) replaces the flux value of central pixel with the median value in the running window and then the maximum filter (we set wavelength step = 120) will replace it with the maximum value. The former smoothes out noisy and the later ignores deeper fluxes that belong to absorption lines (the continuum will be placed in slightly upper or lower locations depending on the values of those parameters). Once the spectrum is filtered, the continuum model will be fitted with a group of splines (we set splines number = 8 and degree = 2) and finally the spectrum is normalized by dividing all the fluxes by the model. One example (Object ``spSpec-54612-2922-626.fit'' in DR7) is as shown in figure \ref{fig:original_wave}.

\subsection{Preprocessing on Spectra with Degraded Resolution: Data Set 2} \label{pre2}

The Chinese Space Station Optical Survey (CSS-OS) is a planned full sky survey operated by the Chinese Space Station Telescope (CSST) \citep{2019ApJ...883..203G}. CSST is a 2-meter space telescope in the same orbit of the China Manned Space Station, which is planned to be launched at the end of 2022. There are three spectroscopic bands covering 255-1000 nm and the resolutions of these bands are all about 200 \citep{2019ApJ...883..203G}.

\begin{figure}[htbp]
    \plotone{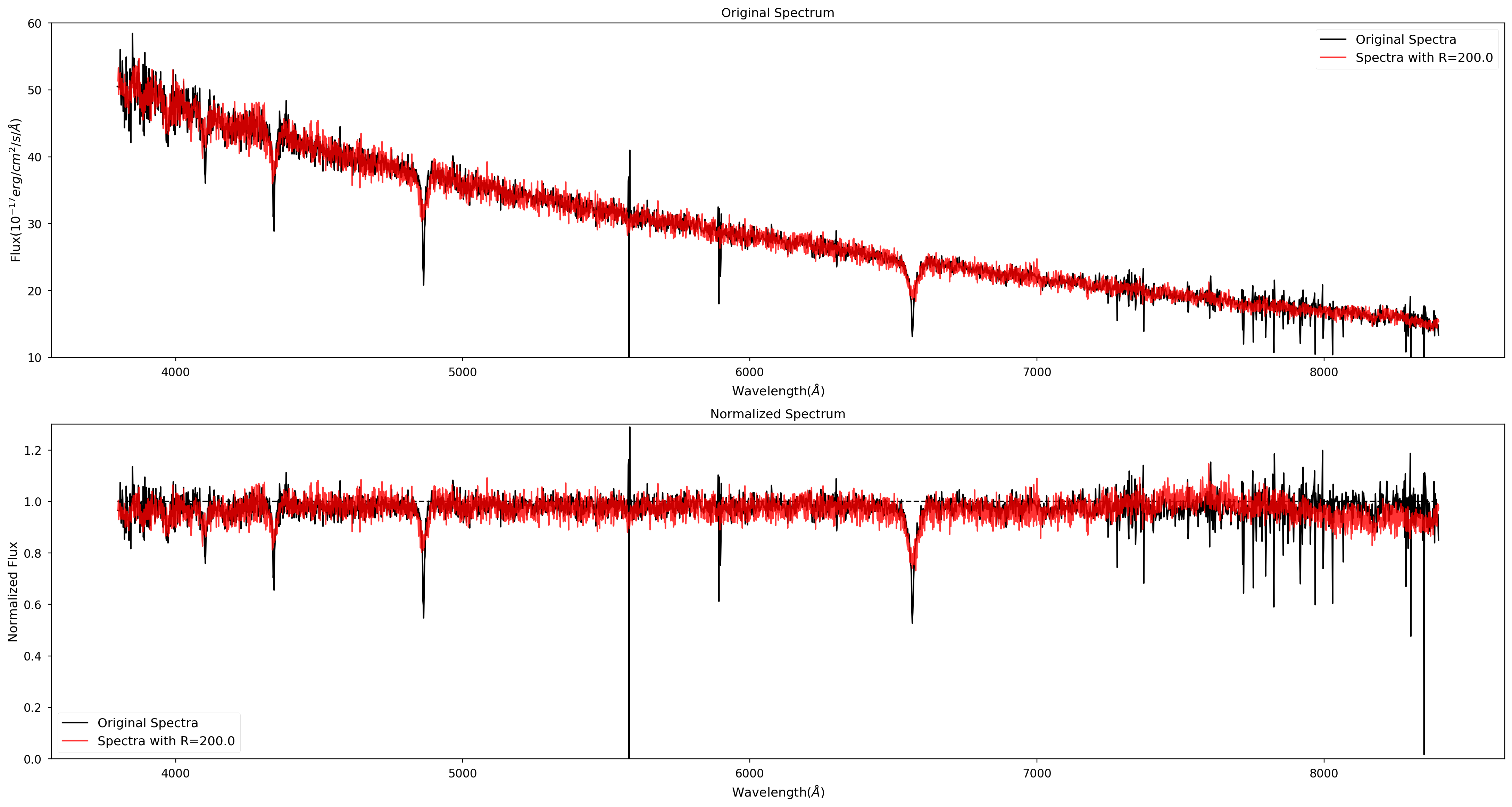}
    \caption{Observed and synthetical spectrum of ``spSpec-52943-1584-513.fit'' in DR7.  \label{fig:wave_200}}
\end{figure}

To figure out whether our deep learning method is practicable under this situation, we plan to test with synthetical spectra. Firstly, we interpolate the original spectra with equal interval of wavelength ($\Delta \lambda =1$ \AA). Then we degrade the resolution of all interpolated spectra to 200 and Gaussian noise ($\sigma$ = flux/30) is added to simulate real spectra observed by CSST. Finally, we normalize them in the same way as mentioned above. Figure \ref{fig:wave_200} shows an example (object ``spSpec-52943-1584-513.fit'' in DR7) of the data reduction.

For all spectra, there are few characteristics of absorption lines and much noise after 7500 \AA\ , so we ignore this portion. Considering actual conditions of all samples, we decide to use fluxes between the range of wavelength from 3834.4 \AA\ to 7500.7 \AA\ (log$\lambda$(\AA) $\in$[3.5837 , 3.8751]), a common region for all spectra, as input of the network.

\section{Method} \label{method}

\subsection{Network} \label{network}

ANN has been widely applied to data analysis in many fields thanks to the vast development of computing power and availability of large data sets. A neural network is a series of algorithms that aims to map the relationships between input data and output labels that we care about. The most important feature of neural networks is that they are adaptive which means they can change or adapt the parameters like weights and biases (see below) within themselves as they learn from continuous training. A deep learning neural network is ``stacked neural networks'' composed of many layers.

\begin{figure}[htbp]
    \epsscale{0.4}
    \plotone{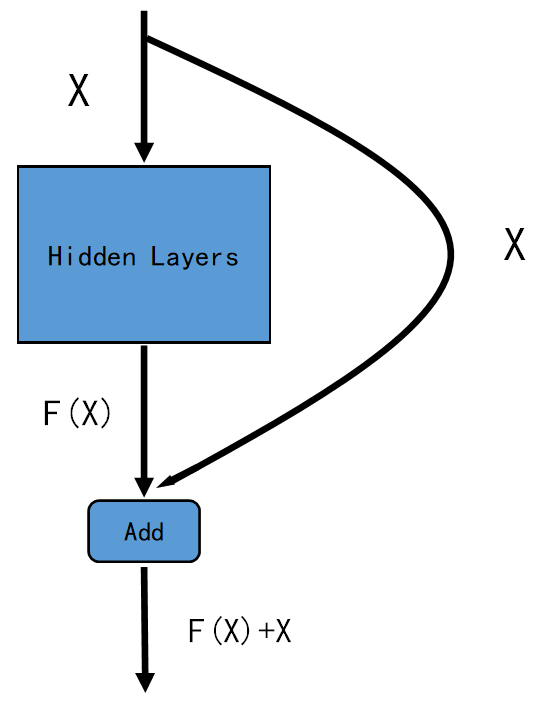}
    \caption{The outline of ResNet block.  \label{fig:resnet_block}}
\end{figure}

\citet{2015arXiv151203385H} introduced Residual Network (ResNet) to solve the problems caused by vanishing or exploding gradient when training very deep neural network. The structure of ResNet block is illustrated in figure \ref{fig:resnet_block}. Output of each block is related to the results of previous hidden layers, as well as the initial input. We adopt this as the basic structure of our network because of its good performance after experiments.

\begin{figure}[htbp]
    \plotone{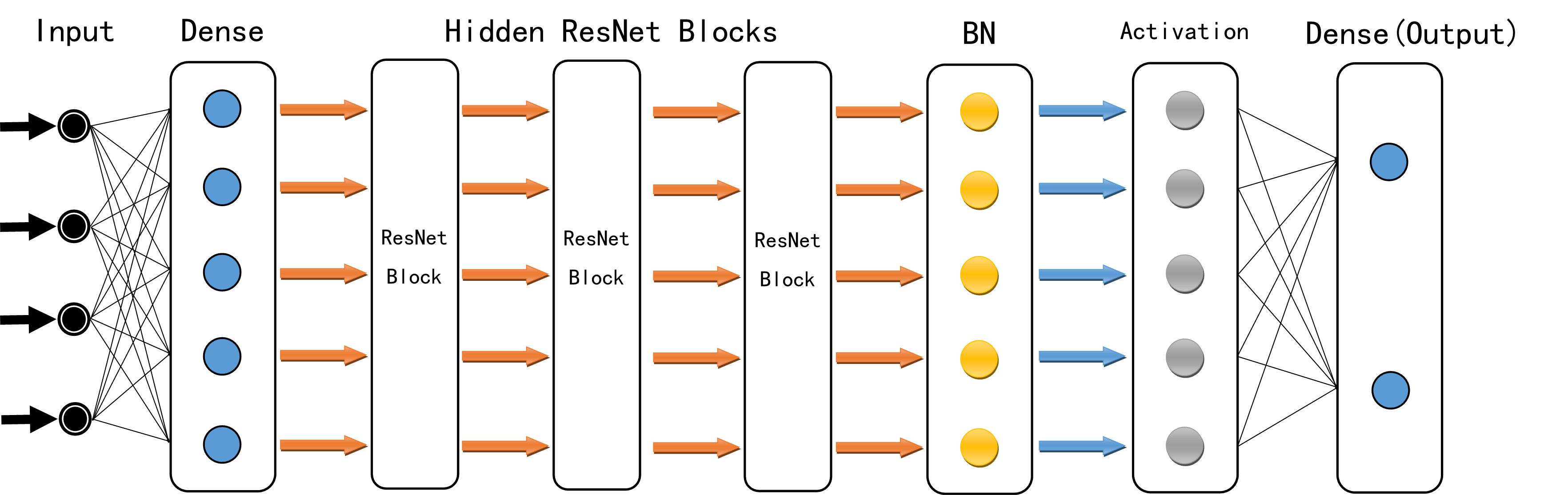}
    \plotone{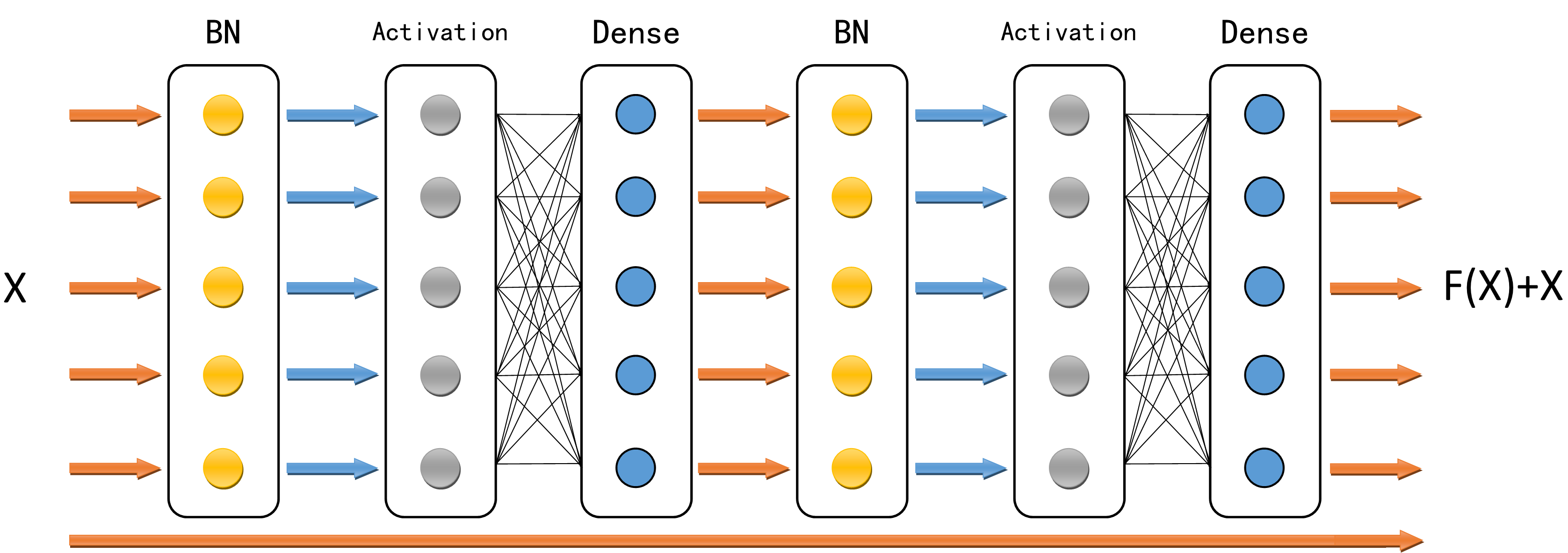}
    \caption{Upper panel: the frame of our network. Lower panel: the details of each ResNet block. \label{fig:network}}
\end{figure}

The frame of our network is displayed in figure \ref{fig:network}. First of all, each flux value of a continuum-normalized spectrum is allocated to a node (black point) of Input layer. All of nodes are fully connected to neurons (blue points) in the next Dense layer, so a neuron will receive all fluxes delivered by the nodes. Output of \textit{n}th neuron $y^{(n)}$ is determined by equation \ref{eq:1} where $x$ is received fluxes, $m$ is the number of flux pixels (nodes) whereas parameters $w^{(n)}$ and $b^{(n)}$ stand for weights and biases of this neuron which are going to be adjusted during training. These outputs are sequentially passed to ResNet blocks.

\begin{equation}
y^{(n)} = \sum \limits ^m_{i=1} w^{(n)}_ix_i +b^{(n)} \label{eq:1}
\end{equation}

Batch Normalization (BN) layer \citep{2015arXiv150203167I} applies a transformation that maintains the mean close to 0 and the standard deviation close to 1 for each neuron channel over training mini-batch (the data are usually divided into several batches and network is successively trained on the data of one batch size at a time rather than the whole data set due to limited computer memory). Use of BN layer will dramatically accelerate the training process and improve performance \citep{2015arXiv150203167I}. Activation layer is used to impart non-linearity to inputs from the last layer to improve performance and stability of the network. The output of each node (gray point) in Activation will be passed to neurons in the next Dense and then neurons will sequentially transmit their outputs calculated with the same mechanism as described in equation \ref{eq:1} to the second sub-block. It is worth nothing that eventual outputs of one ResNet block is obtained by adding the outputs (F(X)) computed from the whole block and the initial inputs (X) as mentioned above.

After processes operated by three ResNet blocks, ``data flow'' is continuously passed to the rest layers. The final layer is still a Dense with only two neurons representing the outlets of \Teff\ and \logg. We instantiate two networks based on two data sets with the frame as shown in figure \ref{fig:network}. For data set 1, the number of nodes in Input is 2915 which is determined by the number of flux pixels. We place 280 neurons in each Dense except the last Dense with two. For data set 2, there are 3666 nodes in Input since we interpolate observed spectra (wavelength interval is 0.0001 in logarithmic form) with the wavelength step 1\AA\ and the number of neurons in Dense is 340. For both cases, we use ``ReLU'' function that returns the maximum between the input and 0 in Activation. Other configurations will be fixed automatically.

\subsection{Training and test} \label{train_test}

\begin{figure}[htbp]
    \plotone{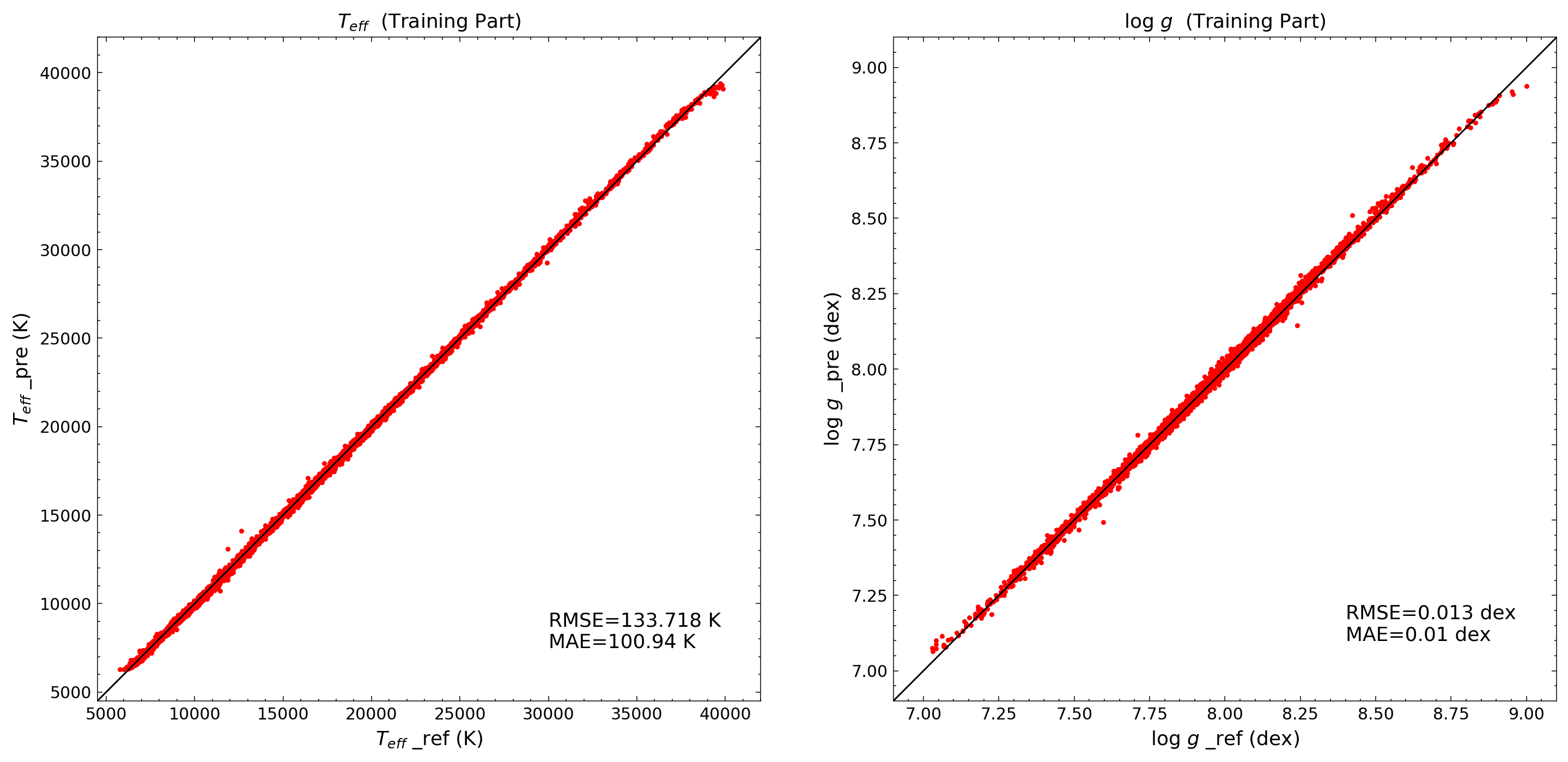}
    \plotone{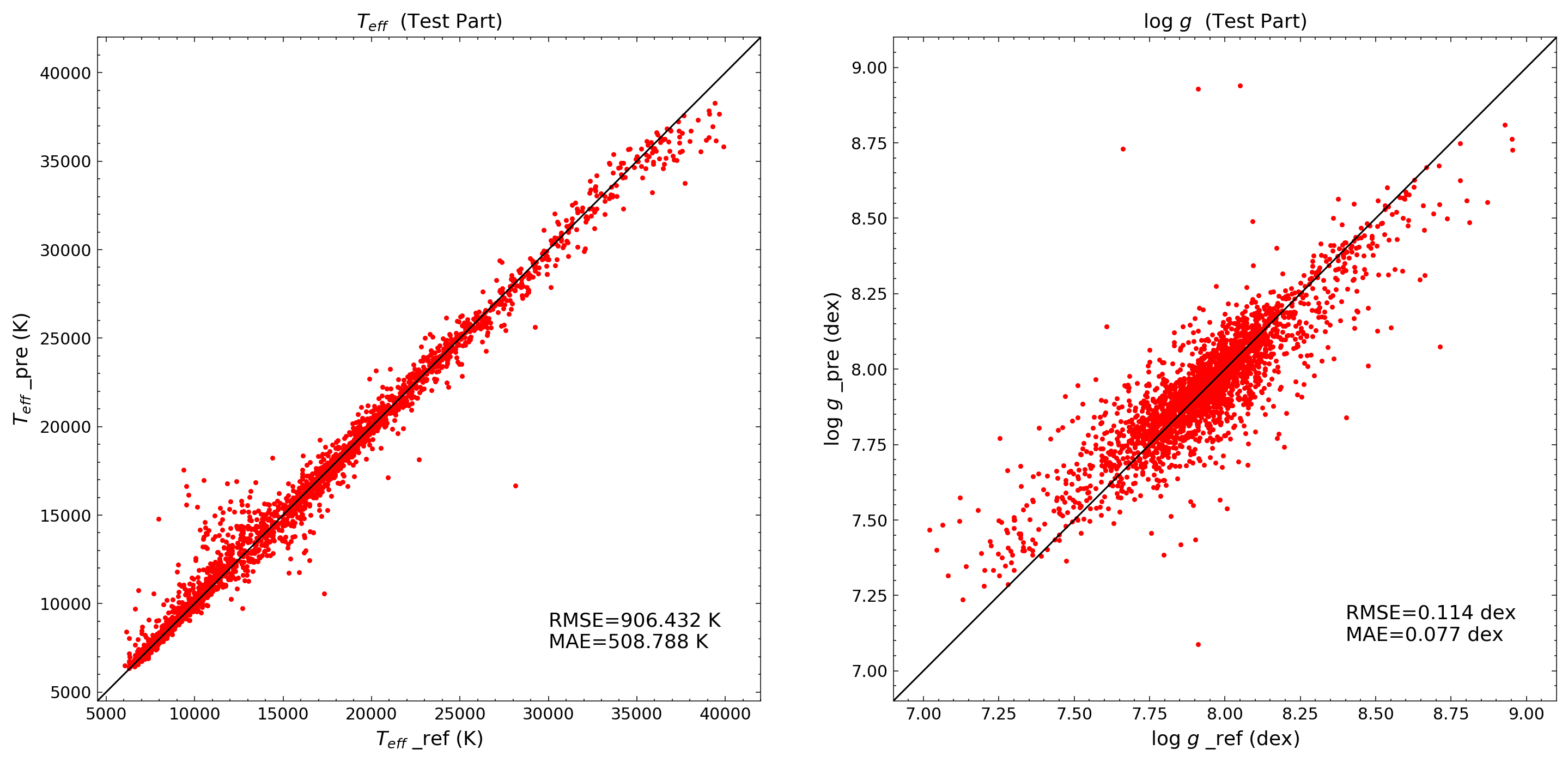}
    \caption{Results of data set 1. Red points: predicted parameters by the network versus reference parameters from WDs catalogs in published papers. Solid line means there is no error. \label{fig:da}}
\end{figure}

\begin{figure}[htbp]
    \plotone{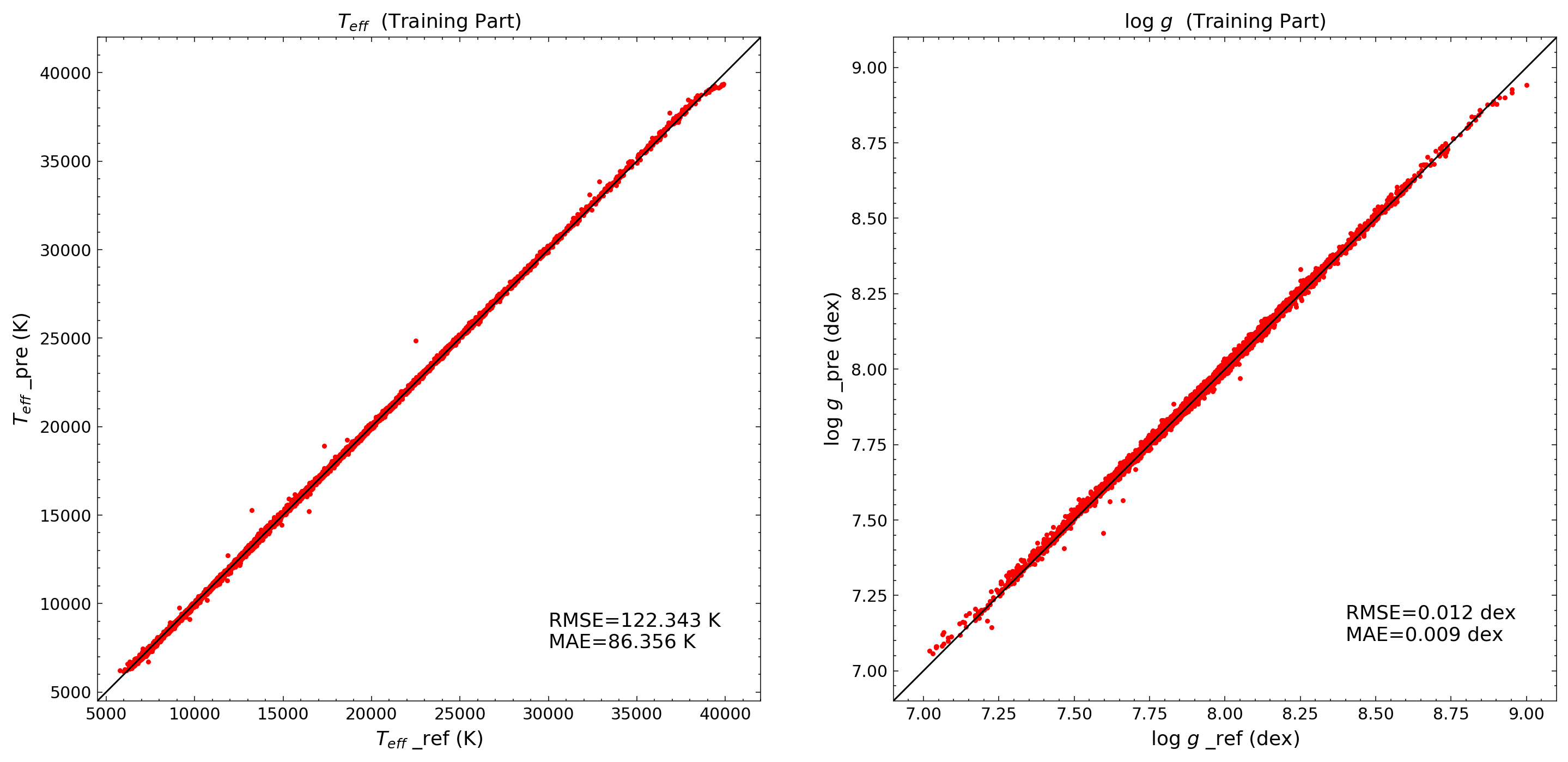}
    \plotone{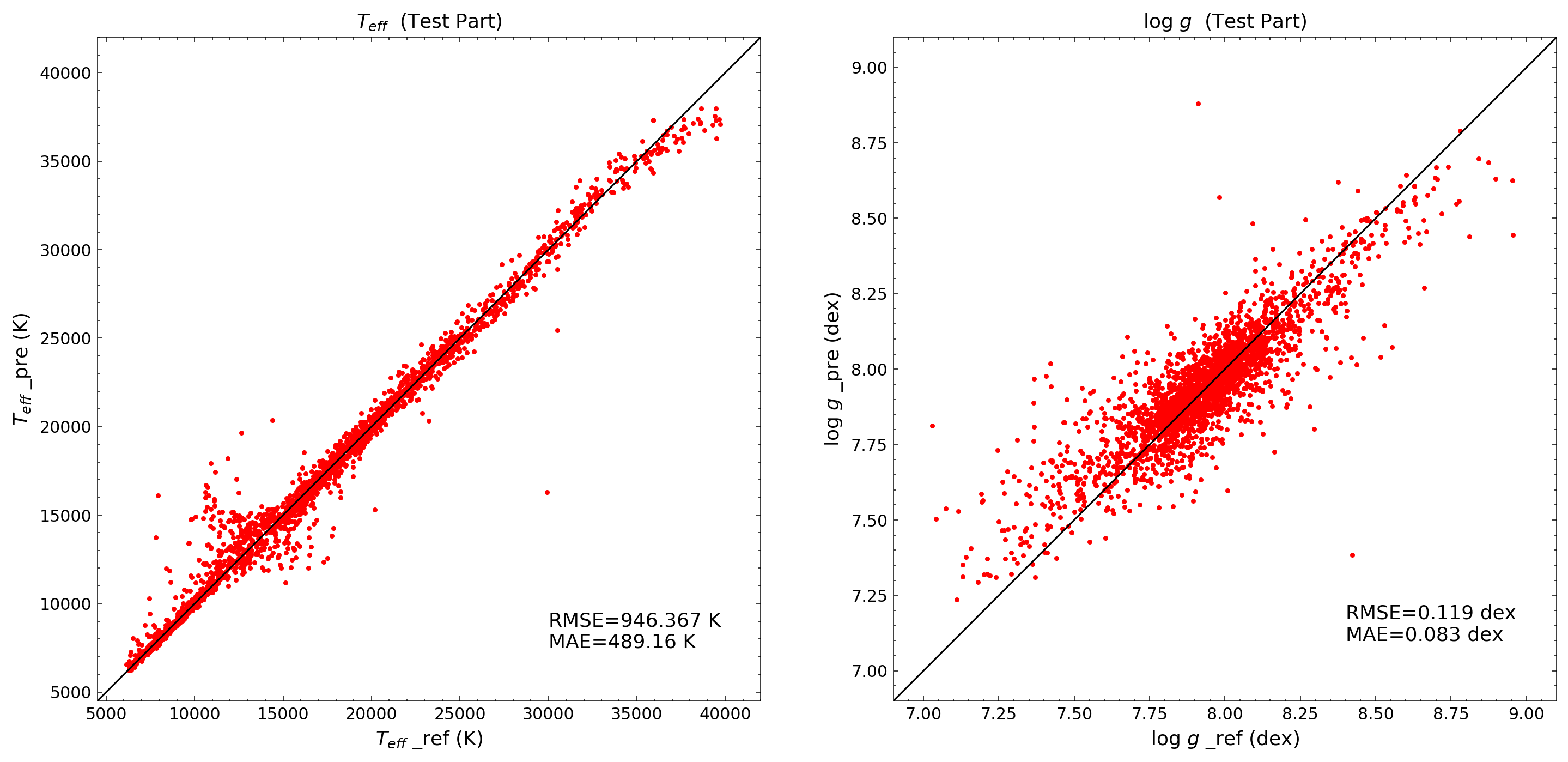}
    \caption{Same as figure \ref{fig:da} but for data set 2.  \label{fig:da_200}}
\end{figure}

We obtain \Teff\ and \logg\ of DAs from the WDs catalogs reported by \citet{2013ApJS..204....5K} and \citet{2015MNRAS.446.4078K, 2016MNRAS.455.3413K} and match these parameters with corresponding spectra. The parameters were determined through theoretical fitting and able to generally describe actual states of the DAs. We adopt these parameters as reference values to construct and verify our networks. After 3D correction, we filter the data with certain criteria and get 8490 samples. Then we preprocess these spectra in two ways yielding two data sets: normalized observed spectra and synthetical spectra with R=200. We split them randomly into two parts: 70\% for training (5943) and 30\% for test (2547). Finally, we use continuum-normalized fluxes as input and normalized \Teff, \logg\ (i.e. minimum is 0 and maximum is 1) as output of networks. The results of each part of every data set are shown in figure \ref{fig:da} and figure \ref{fig:da_200}. We adopt root mean square error (RMSE) and mean absolute error (MAE) (see equation \ref{eq:2}, where $z_i$ is reference value and $\hat{z}_i$ is the parameter estimated by networks) as metrics of performance of the networks.

\begin{equation}
\centering
RMSE = \sqrt{\frac{1}{N}\sum\limits^N_{i=1}(z_i-\hat{z}_i)^2} \ ,\quad
MAE = \frac{1}{N}\sum\limits^N_{i=1}|z_i-\hat{z}_i|
\label{eq:2}
\end{equation}

For training part of data set 1, RMSE and MAE of \Teff\ we get are 133.7 K and 100.9 K, respectively. As for \logg, they are 0.01 dex and 0.01 dex, which means that the network has fully learned properties of training data.

Test part is used to provide an unbiased evaluation on the final model. The tendency of result for \Teff\ generally seems to be fine although there is a little dispersion at \Teff\ $\sim 11000$ K owing to the spectral degeneracy of DAs \citep{2006ApJS..167...40E}, leading to RMSE = 906.4 K and MAE = 508.8 K. The situation for \logg\ is complicated. Since the distribution of \logg\ of our samples is mainly located between 7.0 - 9.0 dex, especially $\sim 8.0$ dex (see figure \ref{fig:all_dist}), the performance is better when \logg\ $\in$[7.75 , 8.25] dex where more data are concentrated. For the side with smaller \logg, predicted values are a little higher than reference ones and for the region with higher \logg, the former is mostly smaller than the later, which results in RMSE =0.11 dex and MAE = 0.08 dex. The performance from test part shows that there is no bad overfitting within the network.

In terms of data set 2, we instantiate the second network with the same frame but different configurations, also resulting in similar consequences (see figure \ref{fig:da_200}). The reason why performances of two networks trained on two data sets with different spectra resolutions are analogous may be because there are strong and wide hydrogen lines rather than very close absorption lines, making different resolutions lead to little effect.

\begin{figure}[htbp]
\plotone{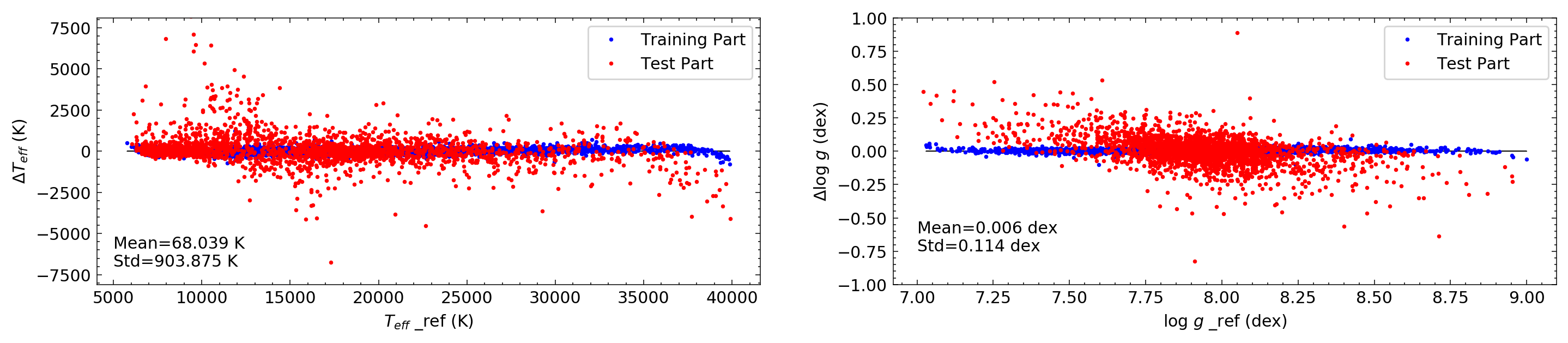}
\plotone{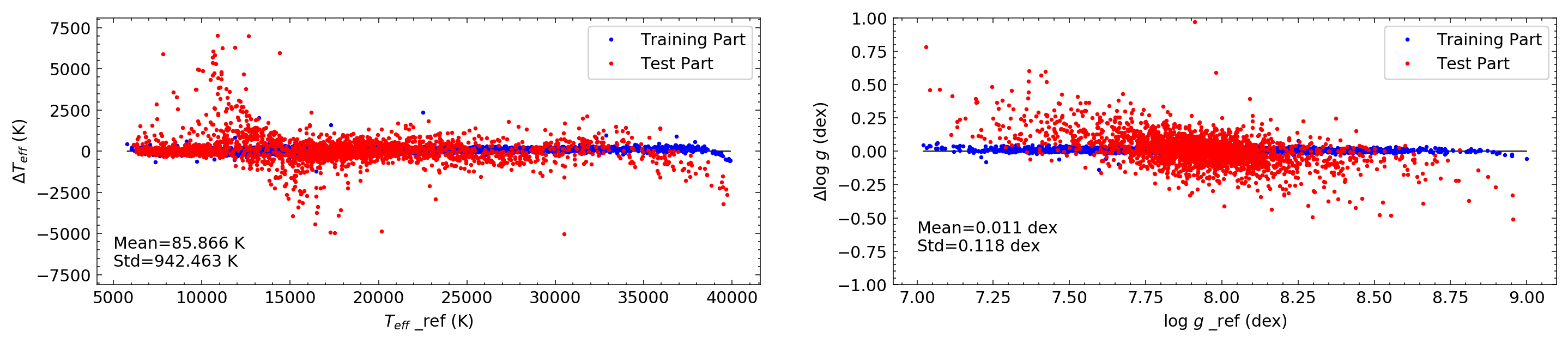}
\caption{Upper panel: residual of parameters of each part for data set 1. Blue points represent training part and red points represent test part. Lower panel: same as the upper but for data set 2. \label{fig:residual} }
\end{figure}

For both data sets, we also calculate the residual ($\Delta = \hat{z} - z$) of parameters for each part. Furthermore, we infer the mean and standard deviation (Std) of the residual for test part, as shown in figure \ref{fig:residual}. 

\section{Conlusion} \label{conclusion}

In this work, we construct a deep learning network based on the ResNet structure. The network is directly trained and tested on the continuum-normalized spectral pixels of DAs spectroscopically confirmed in SDSS DR7, DR10 and DR12 to map the spectra onto atmospheric parameters. The RMSE related to estimated and reference parameters reaches 900 K for \Teff\ and 0.1 dex for \logg. Furthermore, we show that the method is also applicable for the spectra with degraded resolution $\sim 200$.

Compared with the traditional methodology that achieves parameters determination of DAs, our network does not depend on complex theoretical models because it directly use observed normalized fluxes as input to find the matching parameters. The method is feasible for DAs with \Teff\ from 5000 K to 40000 K and \logg\ from 7.0 dex to 9.0 dex.

The existing limitation is that we do not infer error of each estimated parameter due to the running mode of the network. In addition, it only works for DA types currently since it is not easy to recognise and grasp common features from the spectra of other white dwarf types by merely training neural networks.

In the future, we will continue to collect data of WDs and improve the ability of our method to determine physical parameters of the stars more accurately.

\acknowledgments

We thank Qixun Wang for helpful comments and suggestions. This study is supported by the National Natural Science Foundation of China under grant No. 11988101, 11973048, 11927804, 11890694 and National Key R\&D Program of China No. 2019YFA0405502.

Funding for the SDSS and SDSS-II has been provided by the Alfred P. Sloan Foundation, the Participating Institutions, the National Science Foundation, the U.S. Department of Energy, the National Aeronautics and Space Administration, the Japanese Monbukagakusho, the Max Planck Society, and the Higher Education Funding Council for England. The SDSS Web Site is \url{http://www.sdss.org/}.

The SDSS is managed by the Astrophysical Research Consortium for the Participating Institutions. The Participating Institutions are the American Museum of Natural History, Astrophysical Institute Potsdam, University of Basel, University of Cambridge, Case Western Reserve University, University of Chicago, Drexel University, Fermilab, the Institute for Advanced Study, the Japan Participation Group, Johns Hopkins University, the Joint Institute for Nuclear Astrophysics, the Kavli Institute for Particle Astrophysics and Cosmology, the Korean Scientist Group, the Chinese Academy of Sciences (LAMOST), Los Alamos National Laboratory, the Max-Planck-Institute for Astronomy (MPIA), the Max-Planck-Institute for Astrophysics (MPA), New Mexico State University, Ohio State University, University of Pittsburgh, University of Portsmouth, Princeton University, the United States Naval Observatory, and the University of Washington.

Funding for SDSS-III has been provided by the Alfred P. Sloan Foundation, the Participating Institutions, the National Science Foundation, and the U.S. Department of Energy Office of Science. The SDSS-III web site is \break \url{http://www.sdss3.org/}.

SDSS-III is managed by the Astrophysical Research Consortium for the Participating Institutions of the SDSS-III Collaboration including the University of Arizona, the Brazilian Participation Group, Brookhaven National Laboratory, Carnegie Mellon University, University of Florida, the French Participation Group, the German Participation Group, Harvard University, the Instituto de Astrofisica de Canarias, the Michigan State/Notre Dame/JINA Participation Group, Johns Hopkins University, Lawrence Berkeley National Laboratory, Max Planck Institute for Astrophysics, Max Planck Institute for Extraterrestrial Physics, New Mexico State University, New York University, Ohio State University, Pennsylvania State University, University of Portsmouth, Princeton University, the Spanish Participation Group, University of Tokyo, University of Utah, Vanderbilt University, University of Virginia, University of Washington, and Yale University.


\bibliography{sample63}{}
\bibliographystyle{aasjournal}



\end{document}